\documentstyle[12pt,twocolumn]{article}
\def\emi{\end{minipage}}
\def\epi{\end{picture}}
\def\beq{\begin{equation}}
\def\eeq{\end{equation}}
\begin{document} \setcounter{page}{0} \setcounter{equation}{0}
\begin{titlepage}
\vspace*{-1.7cm}
{\normalsize\hfill HD--TVP--94--6}\\
\begin{center}
{
\vspace{1.5cm}
\LARGE\bf
QUARK--GLUON--PLASMA\\
FORMATION\\ \vspace{0.3cm}\par
AT SPS ENERGIES?\footnote{\small Supported in part \hfill by the National
Science
Foundation \hfill under Grant No. PHYS 89-04035}
}
\vspace{1.2cm}\par
{
\large{\bf K. WERNER\footnote{Heisenberg fellow}$^,$\footnote
{\small Email: \hfill werner@dhdmpi5.bitnet,
\hfill werner@hobbit.mpi-hd.mpg.de,
\hfill 28877::werner        }}
\vspace{1.2cm}\par
Institut f\"ur Theoretische Physik, Universit\"at Heidelberg\\
Philosophenweg 19, 69120 Heidelberg, Germany\\ \vspace{0.5cm}\par
and\\ \vspace{0.5cm}\par
Max--Planck--Institut f\"ur Kernphysik, Heidelberg, Germany
}
\end{center}
\vspace{1cm}\par
\begin{quote}
\noindent
By colliding ultrarelativistic ions,
one achieves presently energy
densities close to the critical value, concerning the formation of a
quark-gluon-plasma. This indicates the importance of fluctuations and the
necessity to go beyond
the investigation of average events. Therefore, we introduce a percolation
approach to
model the final stage ($\tau > 1$ fm/c) of ion-ion
collisions, the initial stage being treated by well-established methods,
based on
strings and Pomerons. The percolation approach amounts to finding high density
domains, and treating them as quark-matter  droplets.
In this way, we have a {\bf realistic, microscopic, and Monte--Carlo based
model which allows for
the formation of quark matter.}
We find  that even at SPS
energies large quark-matter droplets are formed -- at a low rate though. In
other
words: large quark-matter droplets are formed due to geometrical fluctuation,
but not
in the average event.
\end{quote}
\vspace{1cm}\par
\vfill
\end{titlepage}
\setcounter{page}{0}
\newpage
\cleardoublepage

Based on extrapolations of present knowledge, one expects that at future
ion-ion
colliders (RHIC, LHC) very high energy densities will be achieved, high enough
to
support the  formation of a new state of matter consisting of deconfined
quarks and
gluons. At present days accelerators for ultrarelativistic  ions (SPS, AGS),
operating  at lower energies, the situation is less clear, because one
reaches energy
densities  just around the critical value. In such a situation, fluctuations
become
crucial, and it is no longer useful to discuss ``average \hbox{events''}.
Therefore,  we
investigate the formation of quark-gluon-matter droplets due to geometrical
fluctuations. This amounts to finding domains of high energy density --
higher than the
average. For this purpose, we introduce a percolation approach for the
final stage,
$\tau> 1$ fm/c, of nucleus-nucleus collisions, the initial
stage being treated by the VENUS string model \cite{wer}.

Straightforward extrapolations of models for (soft) hadron-hadron scattering
turned
out to be quite successful for hadron-nucleus as well as for nucleus-nucleus
scattering. This is, however, only true for very basic observables like the
pion
multiplicity or the transverse energy. The investigation of rare processes
(like $\bar\Lambda$
production) showed the limitations of the ``straightforward extrapolations''
and the
necessity to go beyond. The percolation approach to be discussed later is
such a step
beyond, other mechanisms have been proposed like a hadronic cascade
\cite{ran,ton,sor}
or string fusion \cite{and,paj,sor92}. The percolation appraoch is the first
realistic,
microscopic, and Monte-Carlo based model to allow for the formation of quark
matter.

Let us first discuss the basic ideas of the percolation approach in a
schematic way,
the details and, in particular, the appropriate relativistic formulation
will  be
given later. Consider a snapshot at some fixed time $\tau$. This time  shold be
large
enough ($> 1$ fm/c), so that the ``initial stage''
interactions occuring  in the nucleus-nucleus overlap zone are finished.  At
the given
time $\tau$,  we consider the locations (in $R^3$) of all particles produced
in the
initial stage.  There are, by chance,  regions with high (energy) density and
such
with low density. To be quantitative, we introduce a ``critical energy
density''
$\varepsilon_0$, and look for domains with $\varepsilon >\varepsilon_0$. This
can be
done for example by using a grid and checking the density per cell.
The high density
domains, the connected regions with $\varepsilon>\varepsilon_0$, are referred
to as
quark matter clusters (or droplets). Once they have been formed, these
clusters are
treated macroscopically, characterized by a distribution function for energy
density
$\varepsilon_i(\vec x, \tau)$, momentum $\vec p_i(\vec x, \tau)$, and flavour
$f_i(\vec x,\tau)$,  with $i$ referring to cluster $i$. There are no
constituents
explicitly kept track of, and there is no memory referring to the production
process.

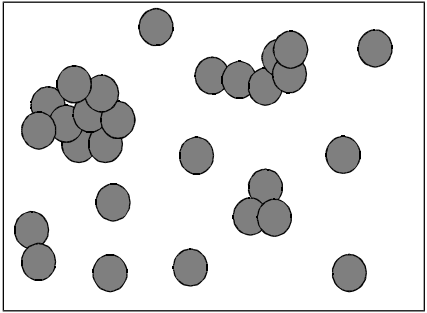
\begin{figure}[t] \unitlength1cm
\begin{minipage}[t]{0.75cm} \begin{picture}(0.75,5)\epi\emi
\begin{minipage}[t]{6.0cm} \begin{picture}(6.0,5)
\epi\par\caption[x]{
Overlapping volumes, representing high energy densities.
}\label{f2}\emi
\hfill\end{figure}

Instead of using a grid, we will use a different but equivalent method: we
assign a
``critical volume'' $V_0$ to all particles - again at given time $\tau$.
Domains of
high energy density correspond to overlapping volumes (see fig. \ref{f2}). So
the task
here is to find connected objects with at least  pairwise overlap of individual
volumes (in fig. \ref{f2}, we find four such objects).  These objects are now
considered as quark matter clusters (or droplets). It should be noted that the
``critical volume'' $V_0$ is not the usual volume  $4\pi R^3_h/3$ of a
hadron, it is
rather defined by the requirement that if hadronic matter is compressed
beyond $V_0$
per hadron, the individual hadrons cease to exist and quark matter is formed.
So $V_0$
is less than $4\pi R^3_h/3$. Rather than the critical volume $V_0$, we
usually use the
critical energy density
\begin{equation}
\varepsilon_0:=\frac{m_h}{V_0},
\end{equation}
with $m_h$ being $(m_\pi+m_\rho)/2$. This critical energy  density
$\varepsilon_0$ (or
equivalently $V_0$) is the crucial  parameter  of our approach. We will
investigate
properties of the system as a function of $\varepsilon_0$, and based on that
we will
try to find the ``realistic value'' of $\varepsilon_0$.

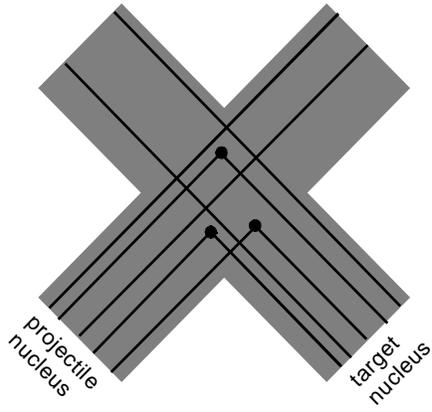
\begin{figure}[t] \unitlength1cm
\begin{minipage}[t]{0.75cm} \begin{picture}(0.75,5.5)\epi\emi
\begin{minipage}[t]{6.0cm} \begin{picture}(6.0,5.5)
\epi\par\caption[x]{
The collision zone (in the t--z plane) of an AA collision. Each dot represents
an origin of a string evolution.
}\label{f3}\emi
\hfill\end{figure}

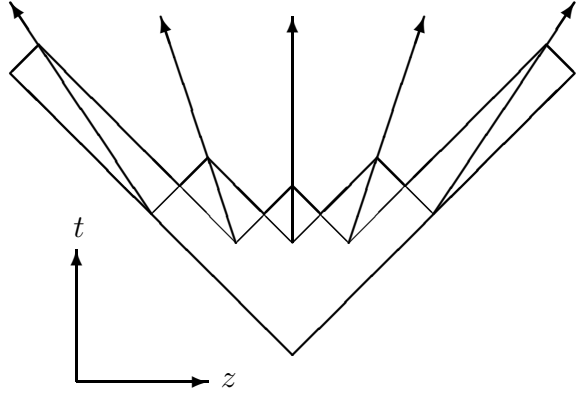
\begin{figure}[t] \unitlength1cm
\begin{minipage}[t]{0.75cm} \begin{picture}(0.75,6.0)\epi\emi
\begin{minipage}[t]{6.0cm} \begin{picture}(6.0,6.0)
\put(0,1.0){\unitlength0.35cm\begin{picture}(8.0,8.0)\thicklines
\put(6.0,0){\unitlength0.25cm\begin{picture}(8.0,8.0)\thicklines
\put(4.5,0){\line(1,1){15}}
\put(4.5,0){\line(-1,1){15}}
\put(4.5,9){\line(1,-1){1.5}}
\put(4.5,9){\line(-1,-1){1.5}}
\put(0,10.5){\line(-1,-1){1.5}}
\put(0,10.5){\line(1,-1){3}}
\put(9,10.5){\line(1,-1){1.5}}
\put(9,10.5){\line(-1,-1){3}}
\put(-9,16.5){\line(-1,-1){1.5}}
\put(-9,16.5){\line(1,-1){7.5}}
\put(18,16.5){\line(1,-1){1.5}}
\put(18,16.5){\line(-1,-1){7.5}}
{\thinlines
\multiput(-3,7.5)(0.25,0.25){1}{\line(-1,1){7.5}}
\multiput(1.5,6.)(0.25,0.25){1}{\line(-1,1){3.0}}
\multiput(4.5,6.)(0.25,0.25){1}{\line(-1,1){1.5}}
\multiput(7.5,6.)(0.25,0.25){1}{\line(-1,1){1.5}}
\multiput(12,7.5)(0.25,0.25){1}{\line(-1,1){1.5}}
\multiput(-3,7.5)(-0.25,0.25){1}{\line(1,1){1.5}}
\multiput(1.5,6.)(-0.25,0.25){1}{\line(1,1){1.5}}
\multiput(4.5,6.)(-0.25,0.25){1}{\line(1,1){1.5}}
\multiput(7.5,6.)(-0.25,0.25){1}{\line(1,1){3.0}}
\multiput(12,7.5)(-0.25,0.25){1}{\line(1,1){7.5}}   }
\put(-3,7.5){\vector(-2,3){7.5}}
\put(1.5,6.){\vector(-1,3){4}}
\put(4.5,6.){\vector( 0,1){12}}
\put(7.5,6.){\vector( 1,3){4}}
\put(12,7.5){\vector( 2,3){7.5}}
\epi
}
\put(1,-1){
\put(0,0){\vector(0,1){5}}
\put(0,0){\vector(1,0){5}}
\put(-.1,5.5){$t$}
\put(5.5,-.2){$z$}
}
\epi}
\epi\par\caption[x]{
Trajectories of string fragments.
}\label{f4}\emi
\hfill\end{figure}

For the first stage  we use the independent string model, to be more precise
the
basic VENUS model without final state interactions \cite{wer}. The elementary
elastic interaction is Pomeron exchange, inelastic scattering and in particular
particle production is treated via the optical  theorem and by using the AGK
cutting
rules. An elementary inelastic process is then represented by a ``cut
Pomeron'',
which amounts to colour exchange and the creation of two strings. In this
formalism,
by using Carlo methods, one can determine projectile and target nucleons which
interact with each other, and how many colour exchanges occur per $NN$
interaction.
In fig. \ref{f3}, we show a typical event: we consider a projection to the
$t$--$z$
plane, nucleon trajectories are represented by straight lines,  interactions
are
represented  by dots. Some nucleons interact (participants), some survive the
interaction zone (spectators). Each dot, representing interaction, is a point
of
string formation, or, in other words, the origin of a string evolution. What
happens
after a string formation point?  We use the standard procedure of classical
relativistic  string dynamics and decays \cite{wer}.  In fig. \ref{f4} a
typical
example of the space-time evolution is shown.  The upper rectangles  represent
produced particles (hadrons and resonances), the arrows indicate particle
trajectories. Remarkable is the strict ordering of the directions, being a
consequence of the covariant string breaking mechanism.

Being able to construct in a first stage, event by event, particle trajectories
defined by their origins in space and the four-momenta, we can proceed to
stage two,
the analysis of energy densities
and cluster formation at fixed time. The question is, what we mean by fixed
time,
which frame we are using. This is going to be discussed in the following.

Crucial for the whole approach is a correlation between the rapidity
\begin{equation}
y:=\frac{1}{2}\ln\frac{E+p_z}{E-p_z}
\end{equation}
and the space-time rapidity
\begin{equation}
\zeta:=\frac{1}{2}\ln\frac{t+z}{t-z},
\end{equation}
with $E$ and $p_z$  being energy and longitudinal momentum  of a particle, $t$
being the time, and $z$ being the longitudinal  coordinate.
The variable $\zeta$ is like an angle, constant $\zeta$ are
straight lines through the origin in the $t$--$z$ plane.
In order
to investigate a correlation between $y$ and $\zeta$, we perform a VENUS
simulation
for central S$+$S collisions at 200 GeV. We measure, for
different times, the average rapidity  $\bar y$ of produced particles in
small cells at
$\vec r_\bot=0$, as a function of $z$. We average over many events.
We observe, as shown in ref. \cite{wer91}, to a very good accuracy
\begin{equation}
\bar y=\zeta,
\end{equation}
with deviations only around $z\approx t$, due to the finite size of the
nuclei. We
may parametrize our findings  as
\begin{equation}
\bar y(t,z)=\left\{\begin{array}{ll}
\zeta^T & \mbox{for} \ z<z^T\\
\zeta(t,z)   & \mbox{for} \ z^T<z<z^P\\
\zeta^P & \mbox{for} \ z>z^P
\end{array}\right.\ ,
\end{equation}
with the boundaries $z^P$ and $z^T$ given as
\beq
z^P(t)=-z^T(t)=\alpha t\ ,
\eeq
with
$\alpha< 1$, and with
\beq
\zeta^P=-\zeta^T=\zeta(t,z^P(t))\ .
\eeq
So we have an ``inner region'' with
Bjorken-type behaviour $\bar y=\zeta$, where all the particle momenta point
back to the
origin, and an ``outer region'' with parallel velocity vectors.
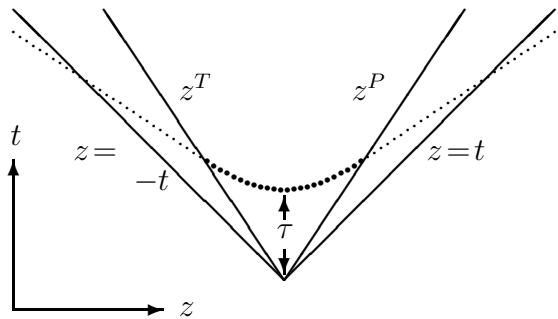
\begin{figure}[t] \unitlength1.0cm
\begin{minipage}[t]{0.75cm} \begin{picture}(0.75,5)\epi\emi
\begin{minipage}[t]{6.0cm} \begin{picture}(6.0,5)
\put(-1.5,1){\unitlength0.40cm\begin{picture}(8.0,8.0)\thicklines
\put(12,0){
\put(0,0){\line(1,1){9}}
\put(0,0){\line(-1,1){9}}
\put(0,0){\line(2,3){6}}
\put(0,0){\line(-2,3){6}}
\put(4.75,4){$z\!=\!t$}
\put(2.25,6){$z^P$}
\put(-7,4){$z\!=$}
\put(-5,3){$-t$}
\put(-3.5,6){$z^T$}
\put(-0.25,1.40){$\tau$}
\put(0,1){\vector(0,-1){0.8}}
\put(0,2){\vector(0,1){0.8}}
\put( -9.000, 8.236){\circle*{0.05}}
\put( -8.778, 8.088){\circle*{0.05}}
\put( -8.556, 7.940){\circle*{0.05}}
\put( -8.333, 7.792){\circle*{0.05}}
\put( -8.111, 7.643){\circle*{0.05}}
\put( -7.889, 7.495){\circle*{0.05}}
\put( -7.667, 7.347){\circle*{0.05}}
\put( -7.444, 7.199){\circle*{0.05}}
\put( -7.222, 7.051){\circle*{0.05}}
\put( -7.000, 6.903){\circle*{0.05}}
\put( -6.778, 6.755){\circle*{0.05}}
\put( -6.556, 6.606){\circle*{0.05}}
\put( -6.333, 6.458){\circle*{0.05}}
\put( -6.111, 6.310){\circle*{0.05}}
\put( -5.889, 6.162){\circle*{0.05}}
\put( -5.667, 6.014){\circle*{0.05}}
\put( -5.444, 5.866){\circle*{0.05}}
\put( -5.222, 5.718){\circle*{0.05}}
\put( -5.000, 5.569){\circle*{0.05}}
\put( -4.778, 5.421){\circle*{0.05}}
\put( -4.556, 5.273){\circle*{0.05}}
\put( -4.333, 5.125){\circle*{0.05}}
\put( -4.111, 4.977){\circle*{0.05}}
\put( -3.889, 4.829){\circle*{0.05}}
\put( -3.667, 4.681){\circle*{0.05}}
\put( -3.444, 4.532){\circle*{0.05}}
\put( -3.222, 4.384){\circle*{0.05}}
\put( -3.000, 4.236){\circle*{0.05}}
\put( -2.778, 4.088){\circle*{0.05}}
\put( -2.556, 3.941){\circle*{0.15}}
\put( -2.333, 3.801){\circle*{0.15}}
\put( -2.111, 3.668){\circle*{0.15}}
\put( -1.889, 3.545){\circle*{0.15}}
\put( -1.667, 3.432){\circle*{0.15}}
\put( -1.444, 3.330){\circle*{0.15}}
\put( -1.222, 3.239){\circle*{0.15}}
\put( -1.000, 3.162){\circle*{0.15}}
\put( -0.778, 3.099){\circle*{0.15}}
\put( -0.556, 3.051){\circle*{0.15}}
\put( -0.333, 3.018){\circle*{0.15}}
\put( -0.111, 3.002){\circle*{0.15}}
\put(  0.111, 3.002){\circle*{0.15}}
\put(  0.333, 3.018){\circle*{0.15}}
\put(  0.556, 3.051){\circle*{0.15}}
\put(  0.778, 3.099){\circle*{0.15}}
\put(  1.000, 3.162){\circle*{0.15}}
\put(  1.222, 3.239){\circle*{0.15}}
\put(  1.444, 3.330){\circle*{0.15}}
\put(  1.667, 3.432){\circle*{0.15}}
\put(  1.889, 3.545){\circle*{0.15}}
\put(  2.111, 3.668){\circle*{0.15}}
\put(  2.333, 3.801){\circle*{0.15}}
\put(  2.556, 3.941){\circle*{0.15}}
\put(  2.778, 4.088){\circle*{0.05}}
\put(  3.000, 4.236){\circle*{0.05}}
\put(  3.222, 4.384){\circle*{0.05}}
\put(  3.444, 4.532){\circle*{0.05}}
\put(  3.667, 4.681){\circle*{0.05}}
\put(  3.889, 4.829){\circle*{0.05}}
\put(  4.111, 4.977){\circle*{0.05}}
\put(  4.333, 5.125){\circle*{0.05}}
\put(  4.556, 5.273){\circle*{0.05}}
\put(  4.778, 5.421){\circle*{0.05}}
\put(  5.000, 5.569){\circle*{0.05}}
\put(  5.222, 5.718){\circle*{0.05}}
\put(  5.444, 5.866){\circle*{0.05}}
\put(  5.667, 6.014){\circle*{0.05}}
\put(  5.889, 6.162){\circle*{0.05}}
\put(  6.111, 6.310){\circle*{0.05}}
\put(  6.333, 6.458){\circle*{0.05}}
\put(  6.556, 6.606){\circle*{0.05}}
\put(  6.778, 6.755){\circle*{0.05}}
\put(  7.000, 6.903){\circle*{0.05}}
\put(  7.222, 7.051){\circle*{0.05}}
\put(  7.444, 7.199){\circle*{0.05}}
\put(  7.667, 7.347){\circle*{0.05}}
\put(  7.889, 7.495){\circle*{0.05}}
\put(  8.111, 7.643){\circle*{0.05}}
\put(  8.333, 7.792){\circle*{0.05}}
\put(  8.556, 7.940){\circle*{0.05}}
\put(  8.778, 8.088){\circle*{0.05}}
\put(  9.000, 8.236){\circle*{0.05}}
}
\put(3,-1){
\put(0,0){\vector(0,1){5}}
\put(0,0){\vector(1,0){5}}
\put(-.1,5.5){$t$}
\put(5.5,-.2){$z$}
}
\epi}
\epi\par\caption[x]{
Space--time evolution of nucleus-nucleus scattering. The dotted line represents
constant proper time (big dots: hyperbola, small dots: tangent).
}\label{f8}\emi
\hfill\end{figure}
Correspondingly, the dotted line in fig. \ref{f8}, a hyperbola in the inner
region
and the tangents at $z^P$ and $z^T$, represent  equal proper time $\tau$ (on
the
average). In this way, obviously, also an average comoving frame is defined.
The
hypersurface defined by the hyperbola/tangents (dotted line) together  with
arbitrary $x_\bot$ is called $\tau$-hypersurface. We are now in a position to
specify
the frame for interactions: we investigate densities (or overlap) at constant
$\tau$,
which means on $\tau$-hypersurfaces.

Having specified the frame and correspondingly the time coordinate $\tau$, we
have to
introduce a useful longitudinal coordinate. We use the ``proper length''
\beq
s:=\int^z_0dz^*, \eeq
with an integration at constant $\tau$, and with $dz^*$ being a longitudinal
length
in the average comoving frame defined by the $\tau$-hypersurface. So we have
\beq
s=\int^z_0\frac{dz'}{\cosh\zeta} \eeq
with $\zeta=\zeta(t,z)=0.5\,\ln(t+z)/(t-z)$ in the inner region and
$\zeta=\zeta^{T/P}$ in the
outer region. In the inner region we have the simple relation
\beq
s=\tau\zeta
\eeq
between the length $s$ and the ``angle'' $\zeta$.

To specify the geometrical properties of particles or clusters, we use the
variables
$\tau$, $s$, and $\vec r_\perp=(r_x,r_y)$. At given $\tau$, a particle or
cluster
is
considered as cylinder in $s,r_x,r_y$ space, with the axis along the $s$-axis.
The object is characterized by a lower and upper value of
$s$, $s_1$ and
$s_2$, and a transverse radius $r_\perp$. We also use
\beq
\bar s:=(s_1+s_2)/2
\eeq
and
\beq
\Delta s:=(s_2-s_1)/2\ .
\eeq

We are now in a shape to construct high density domains, to be considered as
quark
matter cluster. As discussed above, these domains are constructed, as in
percolation
models, by investigating geometrical overlaps of individual objects. There
are two
types of objects: initially, we have only hadrons and resonances, later also
clusters contribute, which have been formed earlier. In any case, at a given
time
$\tau$, the objects are considered cylindric in $(s,r_x,r_y)$-space. Connected
overlapping objects in this space define clusters. Such a
cluster has in general a very irregular shape, which is ``smoothened'' in the
sense
that this complicated shape is replaced by a cylinder of the same volume, the
same
length $s_2-s_1$, and the same center $\bar s$.

Starting at some initial time $\tau_0$ (presently 1 fm/c), we step through time
as $\tau_{i+1}=\tau_i+\delta\tau$, constructing clusters at each time
$\tau_i$. For the time evolution of clusters, we presently assume purely
longitudinal expansion,
\beq
\Delta\zeta(\tau_{i+1})=\Delta\zeta(\tau_i).\eeq

Crucial for our percolation approach is the initialization, i.e.,  the volume
$V_0$
assigned to the hadrons and resonances.
The ``critical''
volume $V_0$, or equivalently the ``critical'' energy
density $\epsilon_0:=m_h/V_0$, is the percolation parameter. To say it again,
the volume $V_0$ is not the usual nucleon volume -- it is the minimum volume
per
particle for hadronic matter to exist. In order to avoid confusion we therefore
prefer to use the critical energy density $\epsilon_0$ as percolation
parameter.
There are two things we are going to do: first, we will investigate, for a
given
reaction, the dynamics as a function of $\epsilon_0$, then we will try to find
a
``realistic'' value of $\epsilon_0$ (denoted as $\epsilon_0^*$), by comparing
with data.

The last and most difficult topic to be discussed is the hadronization. The
power
of our percolation approach is that {\bf any} hadronization scheme can be
plugged
into our approach and tested in a very detailed fashion. This is what we plan
for
the future. Currently, we present a very simple scenario, which can be
implemented
quite easily. Since our cluster expands, the energy density decreases, and
drops
at some stage below $\epsilon_0$. Per definition, we hadronize the cluster at
this
point instantaneously. Since the clusters turn out to have essentially
longitudinal shape ($\Delta s\gg\Delta r_\perp$), we proceed as follows. The
cluster is cut into many short pieces in $s$, all of them having the same
mass $m$
with the requirement of $m$ being around some parameter $m_{\rm seg}$. The
small
clusters then decay isotropically according to phase space \cite{aic}. So this
is essentially the decay of many fireballs at different rapidities.

Our approach is, apart from the relativistic expansion, a typical percolation
problem. In general, one investigates the value for an order parameter $P$  as
a
function of the percolation parameter. For infinite systems, one finds second
order phase transitions, with $P$ being zero in one phase and nonzero in the
other. But also for finite systems a characteristic behaviour survives. In our
case, we expect the following: for large values of the percolation parameter
$\epsilon_0$ (or small $V_0$), we expect the distribution of cluster sizes
being
peaked at small sizes, dropping very fast. For small $\epsilon_0$ (large
$V_0$),
it is most likely to find large clusters, so the distribution will peak at
large
sizes. There should be a transition region, around some ``percolation
transition
value'' $\epsilon_0^{PT}$, with large fluctuations in cluster sizes. The
question
is what is the value of $\epsilon_0^{PT}$, in particular related to the
realistic
value $\epsilon_0^*$ (to be determined).

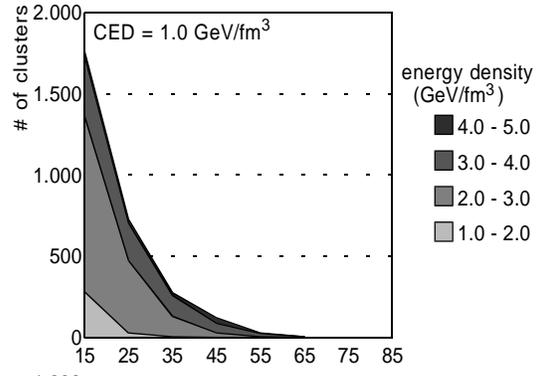
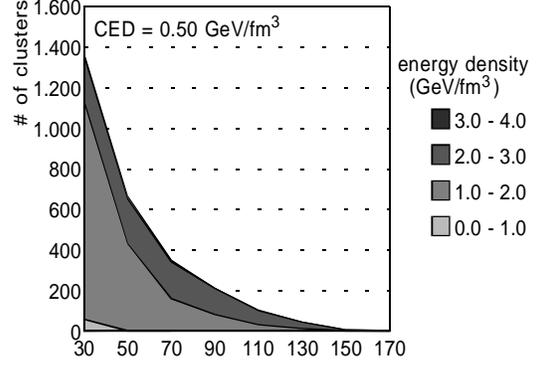
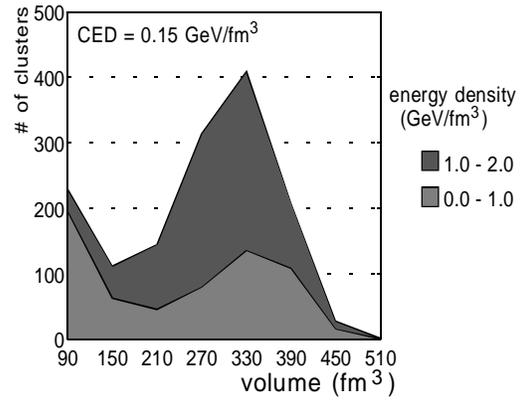
\begin{figure}[t] \unitlength1cm
\begin{minipage}[t]{0.5cm} \begin{picture}(1.0,14.5)\epi\emi
\begin{minipage}[t]{6.0cm} \begin{picture}(6.0,14.5)
\epi\par\caption[x]{
Distribution of cluster volumes for different values of the critical
energy density (CED) $\epsilon_0$.
}\label{f11}\emi
\hfill\end{figure}

In fig. \ref{f11}, we show the results.
We plot the distribution of cluster volumes $V$ for different values of the
critical
energy density (CED) $\varepsilon_0$. The numbers are not normalized, we
show  the
number of clusters per volume bin $\Delta V$, found in 1200 simulations; the
bin
sizes for the three distributions are, from top to bottom: $\Delta V$=10
fm$^3$,
20 fm$^3$, 60 fm$^3$. From the grey-scales, we can also read off how the
clusters are
distributed in energy density. We show only results for $\tau=2$ fm/c, since
volumes
and energy densities turn out to scale in a trivial manner, as $V\sim\tau$ and
$\varepsilon\sim\tau^{-1}$, so the distributions for different $\tau$ look
similar up
to a scale transformation.

We observe exactly what we expected: for $\varepsilon_0=1$ GeV/fm$^3$, the
distribution peaks at small values of $V$, dropping very fast with increasing
volume
$V$. Reducing the CED to 0.50 GeV/fm$^3$, the distribution gets wider,
roughly by a
factor of two. Although, as for $\varepsilon_0=1$ GeV/fm$^3$, very small
clusters are
favoured, the fluctuations are considerably larger. Reducing $\varepsilon_0$
further
to 0.15 GeV/fm$^3$ ($\hat =$ nuclear matter density), we observe a drastic
change in
the distribution, a maximum at large values of $V$ emerged. It is possible to
introduce ``a maximum volume'' $V_{max}$, which is for a longitudinal
expansion given
as
\beq
V_{max}=\tau\Delta y A_\bot,
\eeq
with the rapidity range $\Delta y$ and the transverse area $A_\bot$. For
central
S$+$S at 200 GeV, we obtain for $\tau=2$ fm/c:
\beq
V_{max}\simeq 600 {\rm fm}^3.
\eeq
For $\varepsilon_0=0.15$ GeV/fm$^3$, the tail of the $V$-distribution  almost
reaches
out to this value, and therefore this value of $\varepsilon_0$ is close to the
value $\varepsilon_0^{PT}$ for the percolation transition (if the probability
$P$
for $V_{max}$ to be observed is the order parameter).

It is difficult and not yet conclusive to find a ``realistic'' value
$\varepsilon^*_0$ for $\varepsilon_0$. The best value at present is
$\varepsilon^*_0=1$ GeV/fm$^3$, which provides good results concerning
proton, pion,
and strange particle yields for S$+$S and S$+$Ag scattering. We expect further
confirmation (or not) from investigating particle correlations, because the
value of
$\varepsilon_0$ affects quite strongly ``source radii'' deduced  from
correlation functions. Corresponding work is under progress.

Taking our best value, $\varepsilon_0^*=1$ GeV/fm$^3$, the upper plot of fig.
\ref{f11} represents the realistic world (the lower plots are just mathematical
exercise). Although here it is most likely to produce just small clusters
(mainly
hadrons and resonances), there is nevertheless a reasonable probability to
form big
clusters, with our statistics of 1200 simulations up to 65 fm$^3$. The
question is,
whether such ``miniplasmas'' can be ``isolated'' in event-by-event experiments.

To summarize, we have introduced a percolation approach for the final stage
($\tau > 1$ fm/c) of ultrarelavistic energies. This approach
allows for the formation of quark matter clusters in case of high energy
densities.
We analyzed the distribution of cluster sizes for central S$+$S collisions at
SPS
energies (200 GeV). Although, for a realistic value  of our percolation
parameter
$\varepsilon_0$, it is most likely to form only small clusters (mainly
hadrons and
resonances), there is a reasonable rate to form large clusters. So at  SPS
energies,
it is crucial not to investigate average events, but to pick out events with
large
clusters, which occur due to geometrical fluctuations.


\begin{thebibliography}{999}

\bibitem{wer} K. Werner, Physics Reports 232 (1993) 87-299

\bibitem{ran} H. J. M\"ohring, A. Capella, J. Ranft, J. Tran Thanh Van, C.
Merino, Nucl. Phys. A525 (1991) 493c

\bibitem{ton} V. D. Toneev, A. S. Amelin and K. K. Gudima
preprint GSI-89-52, 1989

\bibitem{sor} H. Sorge, H. St\"ocker and W. Greiner,
Nucl. Phys. A498, 567c (1989)

\bibitem{and} B. Andersson and P. Henning, Nucl. Phys. B335 (1991) 82

\bibitem{paj} M.A. Braun and C. Pajares, Phys. Lett. B287 (1992) 154

\bibitem{sor92} H. Sorge, M. Berenguer, H. St\"ocker, and W. Greiner, Phys.
Lett. B289 (1992) 6

\bibitem{wer91} K. Werner, Nucl. Phys. A525 (1991) 501c

\bibitem{aic} J. Aichelin and K. Werner, Phys. Lett. B300 (1993) 158

\end{thebibliography}
\end{document}